\documentclass[aps,prl,showpacs,twocolumn,amsmath,amssymb,superscriptaddress]{revtex4} 

\usepackage{graphicx}
\usepackage{bm}
\usepackage{mathptmx}
\usepackage{pxfonts}
\usepackage{wasysym} 

\usepackage{Jonasmacros}

\usepackage[colorlinks,citecolor=blue,linkcolor=cyan]{hyperref}
\usepackage[usenames,dvipsnames,svgnames,table]{xcolor}

\begin{document} 
\date{\today} 
\title{Electrical and Thermal Control of Magnetic Exchange Interactions}

\author{Jonas Fransson}
\email{Jonas.Fransson@physics.uu.se}
\affiliation{Department of Physics and Astronomy, Box 516, 75120, Uppsala University, Uppsala, Sweden}

\author{Jie Ren}
\affiliation{Theoretical Division, Los Alamos National Laboratory, Los Alamos, New Mexico 87545, USA} 

\author{Jian-Xin Zhu}
\affiliation{Theoretical Division, Los Alamos National Laboratory, Los Alamos, New Mexico 87545, USA} 
\affiliation{Center for Integrated Nanotechnologies, Los Alamos National Laboratory, Los Alamos, New Mexico 87545, USA} 

\begin{abstract}
We investigate the far-from-equilibrium nature of magnetic anisotropy and exchange interactions between molecular magnets embedded in a tunnel junction. By mapping to an effective spin model, these magnetic interactions can be divided into three types: isotropic Heisenberg, anisotropic Ising, and anisotropic Dzyaloshinski-Moriya contributions, which are attributed to the background non-equilibrium electronic structures. We further demonstrate that both the magnetic self and exchange interactions can be controlled either electrically by gating and tuning voltage bias, or thermally by adjusting temperature bias. We show that the Heisenberg and Ising interactions scale linearly, while the Dzyaloshinski-Moriya interaction scales quadratically, with the molecule-lead coupling strength. The interactions scale linearly with the effective spin-polarizations of the leads and the molecular coherence. Our results pave a way for smart control of magnetic exchange interactions at atomic and molecular levels.
\end{abstract} 

\maketitle

Magnetic interactions is a field of continuously intense activities addressing questions ranging from fundamental physics to technological applications. While control of magnetic interactions is straightforward using magnetic field,  control by the means of electric field presently is an emerging technique. Technological advances such as magnetic memories, magnetic logic gates, and quantum computation, can be envisioned once current controlled magnetic logic circuits have been achieved.

On the one hand, as the technological advances are striving towards the atomic and molecular scale, experiments on magnetic atoms adsorbed onto different surface materials have demonstrated anisotropic effects on spin excitations \cite{hirjibehedin2006,wahl2007,meier2008,balashov2009}, anisotropic Ruderman-Kittel-Kasuya-Yosida (RKKY) interaction \cite{zhou2010}, entanglement of spin excitations and Kondo effect \cite{chen2008,otte2009,pruser2011}, and formation of stable magnetic configurations \cite{khajetoorians2011,loth2012,khajetoorians2013}. Molecular magnets have also been realized in various molecular complexes comprising transition metal atoms \cite{wende2007,fernandez2008,chen2008,chiesa2013,raman2013,fahrendorf2013}, single molecular magnets \cite{carretta2004,cornia2004} and antiferromagnetic rings \cite{slageren2002,carretta2003,troiani2005,carretta2007,wedge2012,candini2010,santini2011}. These experimental advances open new alternatives to design multi-functionalities of nanoscale devices \cite{mannini2010,mannini2009,carretta2006,leuenberger2001,troiani2005,timco2009}.

On the other hand, the theoretical understanding of magnetic interactions at nanocale develops at a fast pace. Recent theoretical advances include phenomenological and microscopic descriptions of spin dynamics \cite{zhang2009,bhattacharjee2012}, non-equilibrium formulation of RKKY interaction \cite{fransson2010}, detailed analysis of exchange interactions in non-collinear magnetic materials \cite{szilva2013}, and magnetic anisotropy in quantum spintronics \cite{misiorny2013}. However, a comprehensive fundamental understanding of the microscopic mechanism of magnetic interactions is still lacking, which hinders us from more flexible control of spin dynamics at nanoscale. 

Here, we uncover the far-from-equilibrium nature of magnetic interactions between molecular magnets embedded between metallic leads. We find that magnetic self and exchange interactions, which are effectively mediated by the electron flow in the system, can be partitioned into isotropic Heisenberg, anisotropic Ising and Dzyaloshinski-Moriya (DM) interactions. The first two interactions scale linearly with the strength of coupling to the leads while the DM interaction scales quadratically.
The interactions, moreover, scale linearly with the effective spin-polarizations of the leads and the molecular coherence. We demonstrate that both the magnitude and the character of the interaction, i.e. ferromagnetic or antiferromagnetic, can be controlled electrically by gating and tuning voltage bias, and thermally by adjusting temperature bias between the leads. Our results for the self interactions reproduce and generalize the results for magnetic anisotropy discussed in Ref.~\onlinecite{misiorny2013}, hence our focus in this paper is on the exchange interactions.

\begin{figure}[b]
\begin{center}
\includegraphics[width=0.99\columnwidth]{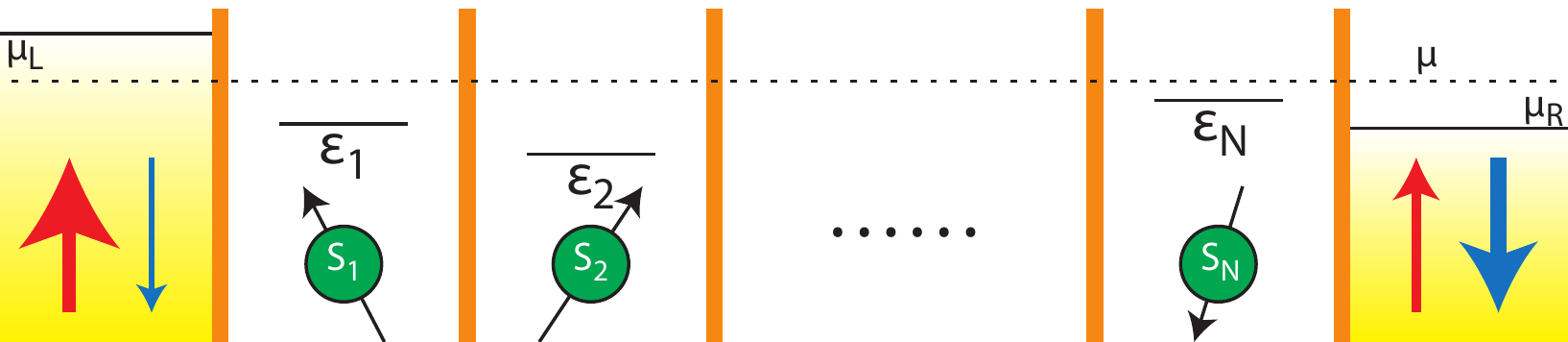}
\end{center}
\caption{(Color online) Sketch of magnetic molecules embedded in a junction between magnetic leads. Electrons may tunnel between the electrodes and the localized levels $\dote{n}$ and between the levels. An electron residing in level $n$ interacts with the localized spin moment $S_n$. The set-up may be achieved by, e.g., stacking molecules on top of each other on a surface, or constructing a chain on an insulating surface, using scanning tunneling microscopy techniques.}
\label{fig-1}
\end{figure}

We model the magnetic molecule $n$ by a spin moment $\bfS_n$, which is coupled to a single level $\leade{n}$ via exchange $\Hamil_{int}=\sum_nv_n\bfs_n\cdot\bfS_n$, see Fig.~\ref{fig-1}. Here, $\bfs_n=\sum_{\sigma\sigma'}\cdagger{n}\bfsigma_{\sigma\sigma'}\cs{n\sigma'}/2$ represents the delocalized electron spin, where $\cdagger{n}$ ($\cc{n}$) denote the electron creation (annihilation) in the single level of the $n$th molecule, whereas $v_n$ is the coupling strength, and $\bfsigma$ is the vector of Pauli matrices. The molecular complex is represented by $\Hamil_M=\sum_{n\sigma}[\dote{n\sigma}c^\dagger_{n\sigma}\cs{n\sigma}+\mathcal{T}_c(c^\dagger_{n\sigma}\cs{n+1\sigma}+H.c.)]+\Hamil_{int}$, where $\calT_c$ denotes the tunneling rate between adjacent molecules \cite{note1}. The molecules are coupled to the electrodes with rate $\mathcal{T}_\chi$, $\chi=L,R$, where $L\ (R)$ denotes the left (right) lead. The leads are specified by their respective chemical potential $\mu_\chi$ and temperature $T_\chi$, and we consider constant voltage and thermal bias. The full system is represented by the Hamiltonian
\begin{align}
\Hamil=&
	\Hamil_L+\Hamil_R+\Hamil_T+\Hamil_M.
	\label{eq:eq1}
\end{align}
Here, $\Hamil_\chi=\sum_{\bfk\sigma}(\leade{\bfk}-\mu_\chi)\cdagger{\bfk}\cc{\bfk}$ represents the Hamiltonian for the lead $\chi$, where $\cdagger{\bfk}$ ($\cc{\bfk}$) creates (annihilates) an electron in a lead with energy $\leade{\bfk}$, momentum $\bfk$, and spin $\sigma=\up,\down$, and we shall use $\bfk=\bfp\ (\bfq)$ for the left (right) lead. The tunneling Hamiltonian $\Hamil_T=\Hamil_{TL}+\Hamil_{TR}$, where $\Hamil_{TL}=\mathcal{T}_L\sum_{\bfp\sigma}\cdagger{\bfp}\cs{1\sigma}+H.c.$, and analogously for the right interaction, assuming that the spin is conserved in the tunneling process.
The model we use pertains to, e.g., paramagnetic M-phthalocyanine (MPc) and M-porphyrine molecules \cite{wende2007,chen2008,coronado2004}, where M denotes a transition metal element (Cu, Fe, Ni, Co, Mn), and similar structures where the magnetic moment is carried by the transition metal $d$-orbitals which are weakly interacting with the delocalized $s$- and $p$-orbitals that carry the charge conduction.

The local interactions between the spin moment $\bfS_n$ and electrons in level $\leade{n}$ give rise to a contribution $\delta{\cal S}$ to effective spin action ${\cal S}_{\rm eff}$ \cite{zhu2004,fransson2008,bhattacharjee2012}, given by
\begin{align}
\delta{\cal S}=&
	\frac{1}{e}
	\sum_{mn}
	\int
		[		
			\boldsymbol{\epsilon}_m\bfj_{mn}(t,t')
			+{\bfS_{m}}(t)\cdot\mathbb{J}_{mn}(t,t')
		]
		\cdot\bfS_n(t')
	dt'dt,
\label{eq-action}
\end{align}
The contribution $\boldsymbol{\epsilon}_m\bfj_{mn}=ie\boldsymbol{\epsilon}_mJ_n\theta(t-t')\av{\com{\bfs_m^{(0)}(t)}{\bfs_n(t')}}$ provides the magnetic field exerted on the local spin moment due to electron flow. Here, $\boldsymbol{\epsilon}_m=\diag{\varepsilon_{m\up}\ \varepsilon_{m\down}}{}$ and $\bfs^{(0)}_m=\sum_{\sigma\sigma'}\cdagger{m}\sigma^0_{\sigma\sigma'}\cs{m\sigma'}/2=\sum_\sigma\cdagger{m}\cc{m}/2$ is the charge, where $\sigma^0$ is the identity matrix. The current $\mathbb{J}_{mn}=iev_mv_n\theta(t-t')\av{\com{\bfs_m(t)}{\bfs_n(t')}}$ carries the magnetic anisotropy and exchange interactions \emph{between} the local magnetic moments $\bfS_m$. As the first contribution in Eq. (\ref{eq-action}) was discussed in \cite{misiorny2013}, our primary focus will be on the second.

The self interaction $\mathbb{J}_{mm}$ defines the anisotropy field acting on the local spin moment $\bfS_m$, while $\mathbb{J}_{mn}$ mediate the exchange interaction between two different spin moment $\bfS_m$ and $\bfS_n$. For small coupling $v_m$ we can neglect the back-action from the localized spins on the electrons. In the stationary regime we can therefore express the current $\mathbb{J}_{mn}$ in energy space as
\begin{align}
\mathbb{J}_{mn}(\omega)=&
	\frac{e}{4}v_mv_n
	\int
		\frac{1}
		{\omega-\dote{}+\dote{}'+i\delta}
		{\rm sp}\ \bfsigma
		\Bigl(
			\bfG_{mn}^<(\dote{})\bfsigma\bfG_{nm}^>(\dote{}')
\nonumber\\&
			-\bfG_{mn}^>(\dote{})\bfsigma\bfG_{nm}^<(\dote{}')
		\Bigr)
	\frac{d\dote{}}{2\pi}\frac{d\dote{}'}{2\pi}.
\end{align}
Here, $\bfG^{</>}_{mn}$ is the lesser/greater (spin space matrix) Green function (GF) for propagation of an electron from molecule $n$ to $m$. ${\rm sp}$ is the spin space trace and the products ${\rm sp}[\bfsigma\bfG_{mn}][\bfsigma\bfG_{nm}]$ are dyads defined as $\bfa{\bf b}=a_ib_j\hat{\bf i}\hat{\bf j}$ such that $\mathbb{J}_{mn}$ constitutes a tensorial quantity.

The electron GF $\bfG_{mn}$ can always be partitioned into charge and magnetic components, $g_{mn}^{(0)}$ and $\bfg_{mn}^{(1)}$, according to $\bfG_{mn}=g_{mn}^{(0)}\sigma^0+\bfg_{mn}^{(1)}\cdot\bfsigma$. In terms of this notion it is straightforward to see that the localized molecular spins in Eq.~(\ref{eq:eq1}) can be mapped into an effective Hamiltonian $\Hamil_S$ corresponding to the interaction $\int\bfS_m\cdot\mathbb{J}_{mn}\cdot\bfS_ndtdt'/e$. This effective spin interaction model can be written as
\begin{align}
\Hamil_S=&
	\sum_{mn}
		\bfS_m\cdot
		\Bigl(
			J_{mn}\bfS_n
			+\bfI_{mn}\cdot\bfS_n
			+\bfD_{mn}\times\bfS_n
		\Bigr),
\label{eq-Hs}
\end{align}
where the three contributions in the above model describe Heisenberg, Ising, and DM interactions, respectively, given by the $\omega\rightarrow0$ limit of the integrals
\begin{subequations}
\label{eq-jmn}
\begin{align}
J_{mn}(\omega)=&
	\frac{1}{2}v_mv_n
	\fint
		\frac{1}{\omega-\dote{}+\dote{}'}
	\Bigl(
		g^{(0)<}_{mn}(\dote{})
		g^{(0)>}_{nm}(\dote{}')
\nonumber\\&
		-
		g^{(0)>}_{mn}(\dote{})
		g^{(0)<}_{nm}(\dote{}')
		-
		\bfg^{(1)<}_{mn}(\dote{})
		\cdot
		\bfg^{(1)>}_{nm}(\dote{}')
\nonumber\\&
		+
		\bfg^{(1)>}_{mn}(\dote{})
		\cdot
		\bfg^{(1)<}_{nm}(\dote{}')
		\Bigr)
	\frac{d\dote{}}{2\pi}
	\frac{d\dote{}'}{2\pi}
	,
\label{eq-jH}
\\
\bfI_{mn}(\omega)=&
	\frac{1}{2}v_mv_n
	\fint
		\frac{1}{\omega-\dote{}+\dote{}'}
		\Bigl(
		\bfg^{(1)<}_{mn}(\dote{})
		\bfg^{(1)>}_{nm}(\dote{}')
\nonumber\\&
		-
		\bfg^{(1)>}_{mn}(\dote{})
		\bfg^{(1)<}_{nm}(\dote{}')
		+
		\bfg^{(1)<}_{nm}(\dote{}')
		\bfg^{(1)>}_{mn}(\dote{})
\nonumber\\&
		-
		\bfg^{(1)>}_{nm}(\dote{}')
		\bfg^{(1)<}_{mn}(\dote{})
	\Bigr)
	\frac{d\dote{}}{2\pi}
	\frac{d\dote{}'}{2\pi}
	,
\label{eq-jI}
\\
\bfD_{mn}(\omega)=&
	\frac{1}{4}v_mv_n
	\fint
		\Bigl(
		g^{(0)<}_{mn}(\dote{}+\omega)
		\bfg^{(1)>}_{nm}(\dote{})
\nonumber\\&
		-
		g^{(0)>}_{mn}(\dote{}+\omega)
		\bfg^{(1)<}_{nm}(\dote{})
		-
		\bfg^{(1)<}_{mn}(\dote{}+\omega)
		g^{(0)>}_{nm}(\dote{})
\nonumber\\&
		+
		\bfg^{(1)>}_{mn}(\dote{}+\omega)
		g^{(0)<}_{nm}(\dote{})
		\Bigr)
	\frac{d\dote{}}{2\pi}
	,
\label{eq-jDM}
\end{align}
\end{subequations}
where $\fint$ denotes the Cauchy principal value. Negative (positive) parameters $J_{mn}$, $\bfI_{mn}$, and $\bfD_{mn}$ correspond to ferromagnetic (antiferromagnetic) interactions.

We notice here, for instance, that the Heisenberg like interaction is finite regardless of the spin-polarization in the molecules, while the Ising and DM like interactions are finite only under spin-polarized conditions. It may also be noticed that the Ising like interaction contributes to the uniaxial anisotropy \cite{IsingE} whereas the DM like interaction provides a transverse anisotropy component.

The expressions for the Heisenberg, Ising, and DM self and exchange interactions given in Eq. (\ref{eq-jmn}) constitute a very general result since they provide the spin-interactions far from equilibrium, as well as in equilibrium, both under electric and thermal fields. The expressions can, moreover, be employed in materials calculations by interpreting the GFs $\bfG_{mn}$ in terms of real space distributions of the electronic structure. In the present context, we shall go deeper into a discussion of their properties in coupled magnetic molecules.

Under equilibrium conditions (vanishing voltage and thermal biases), we can employ the fluctuation-dissipation theorem through the relation $\bfG^{</>}_{mn}(\omega)=(\pm i)f(\pm\omega)[-2\im\bfG^r_{mn}(\omega)]$, where $f(\omega)$ is the Fermi-Dirac distribution function at the (electro-) chemical potential $\mu$. We define $g^{(0)}_{mn}=\sum_\sigma G_{mn\sigma}/2$ and $\bfg^{(1)}_{mn}=\hat{\bf z}\sum_\sigma\sigma^z_{\sigma\sigma}G_{mn\sigma}/2$, for a simple collinear spin-polarized structure. Inserting into Eq.~(\ref{eq-jH}) and using the Kramers-Kr\"onig relations we obtain $J_{mn}=v_mv_n \im\sum_\sigma\int f(\dote{})G^r_{mn\sigma}(\dote{})G^r_{nm\bar\sigma}(\dote{})d\dote{}/4\pi$, which is in agreement with previous results \cite{blackman1969,antropov1995,note2}.

Under non-equilibrium conditions we use the general relation $\bfG^{</>}(\omega)=\bfG^r(\omega)\bfSigma^{</>}(\omega)\bfG^a(\omega)$, where the self-energy $\bfSigma^{</>}$ is given by the couplings $\Gamma^\chi_\sigma$. Reducing the set-up to a molecular dimer and neglecting the back-action from the localized spins, we can write the GF
\begin{align}
\bfG^r_\sigma(\omega)=&
	\frac{1}{2\Omega_\sigma}
	\sum_{s=\pm1}
		\frac{\Omega_\sigma\sigma^0+2s\mathcal{T}_c\sigma^x+s(\Delta_\sigma-i\gamma_\sigma/2)\sigma^z}{\omega-E_{\sigma s}}.
\end{align}
Here, $E_{\sigma\pm}=(\leade{1}+\leade{2}\pm\Omega_\sigma-i\Gamma_\sigma/2)/2$, $\Omega_\sigma^2=(\Delta_\sigma-i\gamma_\sigma/2)^2+4\mathcal{T}_c^2$, $\Delta_\sigma=\leade{1}-\leade{2}$, $\Gamma_\sigma=\sum_\chi\Gamma^\chi_\sigma$, and $\gamma_\sigma=\Gamma^L_\sigma-\Gamma^R_\sigma$. The resonance $E_{\sigma+}$ ($E_{\sigma-}$) signifies the orbital with the highest (lowest) energy, and $\Gamma^\chi_\sigma=2\pi\sum_{\bfk\in\chi}\mathcal{T}^2_\chi\rho^\chi_{\mathbf{k}\sigma}$ denotes the coupling to the lead $\chi=L,R$, in terms of the density of electron states (DOS) $\rho^\chi_{\mathbf{k}\sigma}$. The spin-polarization in the leads is parametrized within a Stoner picture using $p_\chi\in[-1,1]$ and $\Gamma^\chi_\sigma=\Gamma^\chi(1+\sigma^z_{\sigma\sigma}p_\chi)/2$ such that $\Gamma^\chi=\sum_\sigma\Gamma^\chi_\sigma$ and $\Gamma=\sum_\chi\Gamma^\chi$.

\begin{figure}[t]
\begin{center}
\includegraphics[width=0.99\columnwidth]{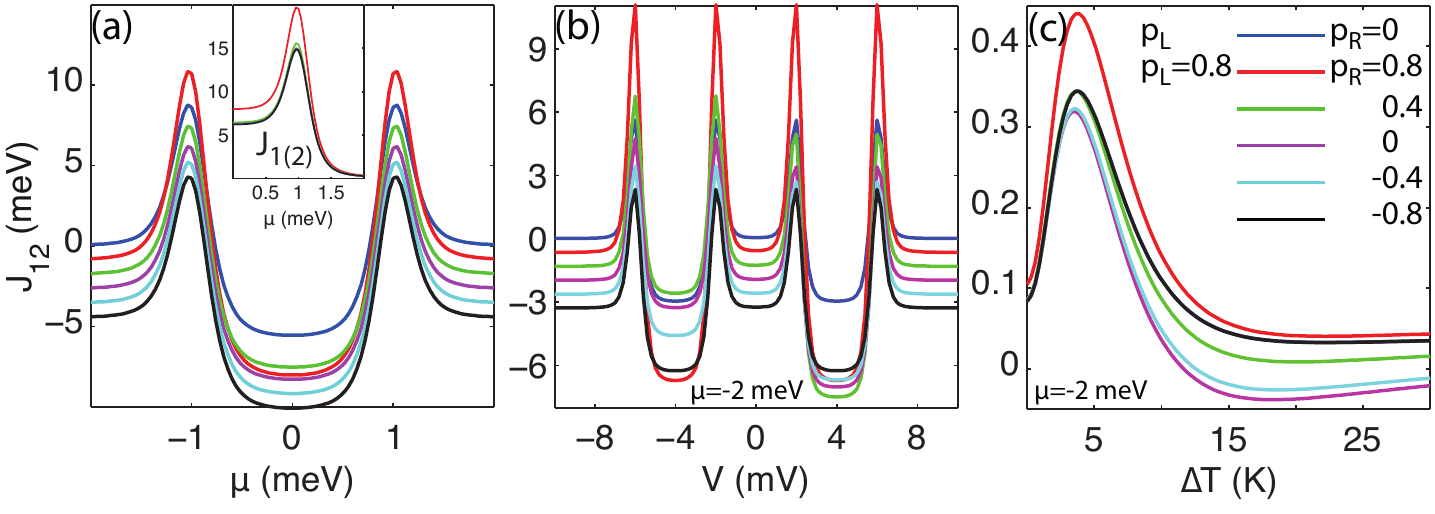}
\end{center}
\caption{(Color online) Heisenberg exchange $J_{12}$ as function of (a) chemical potential $\mu$, (b) bias voltage $V$, and (c) temperature difference $\Delta T=T_R-T_L$. The plots in (a) and (b) are off-set for clarity, the system is gated ($\mu=-2$ meV) in (b), (c), while the colors refer to different spin-polarizations $(p_L,p_R)$ in the leads. The inset in (a) shows the electronically induced anisotropy fields $J_{1(2)}$ acting on the individual spins. Here, $\dote{0}=0$, $\mathcal{T}_c=3\Gamma=v_1/5=v_2/5=1$ meV, and $T_L=1$ K.}
\label{fig-2}
\end{figure}

For the transparency of mathematical formulation, we assume equivalent molecules such that $\leade{n}=\dote{0}$ and symmetric couplings $\Gamma^\chi_\sigma=\Gamma_\sigma/2$, retaining spin-polarization in the leads. The Heisenberg exchange $J_{mn}$ ($m\neq n$) then becomes
\begin{align}
J_{mn}=&
	-\frac{\mathcal{T}_c^2}{8\pi}v_mv_n
	\sum_\sigma\Gamma_\sigma
	\fint
			\frac{
				f_L(\dote{})+f_R(\dote{})
			}
			{
				|\dote{}-E_{\sigma+}|^2
				|\dote{}-E_{\sigma-}|^2
			}
\nonumber\\&\times
			(\dote{}-\dote{0})
			\frac{(\dote{}-\dote{0})^2-\mathcal{T}_c^2-(\Gamma_{\bar\sigma}/4)^2}
			{
				|\dote{}-E_{\bar\sigma+}|^2
				|\dote{}-E_{\bar\sigma-}|^2
			}
	d\dote{}.
\label{eq-Jdimer}
\end{align}
We notice that the Heisenberg exchange depends on the electronic occupations ($\propto~f_L+f_R$) of the leads and scales linearly with $\Gamma$. The expression, moreover, indicates that there is a finite exchange interaction between the localized spins whenever the chemical potential $\mu_\chi$ lies within the energy range of the molecular orbitals, that is, $(\mu_\chi-\dote{0})^2\leq \mathcal{T}_c^2+(\Gamma_\sigma/4)^2$. This result is demonstrated in  Fig.~\ref{fig-2}(a), which shows the equilibrium exchange as function of $\mu_\chi=\mu$ for different spin-polarizations $p_L$ and $p_R$. The exchange, which peaks at the orbital resonances $E_{\sigma\pm}$, is anti-ferromagnetic below $E_{\sigma-}$ (above $E_{\sigma+}$) and ferromagnetic between the resonances, which is a typical behavior for superexchange. This behavior can be controlled by means of gating or tuning voltage bias, see Fig.~\ref{fig-2}(b) where the system is gated ($\mu-\dote{0}=-2$) and driven with a finite voltage bias. Experimental values of antiferromagnetic (Heisenberg) exchange between, e.g., MPc have been reported in the range between 0.5 | 20 meV \cite{chen2008,coronado2004}, and our results are well within this regime for realistic parameters of the model.

From Fig.~\ref{fig-2}(a) and \ref{fig-2}(b), it is clear that the equilibrium and non-equilibrium responses on the spin-polarization in the leads are quite different. While the exchange depends only weakly on $(p_L,p_R)$ in equilibrium, the ferromagnetic regimes change dramatically under non-equilibrium conditions. Current flowing from stronger to weaker spin-polarization generates a stronger ferromagnetic exchange while it becomes weaker when the current flows in the opposite direction. 

Varying the temperature and/or introducing a thermal bias $\Delta T=T_R-T_L$ provides an alternative route to control the exchange. The thermal broadening of the electronic density in the leads effectively makes it (partially) resonant with the molecular orbitals. The plots in Fig.~\ref{fig-2}(c) shows the dependence on a thermal bias for different $(p_L,p_R)$. The initial peak is related to the fact that the lower orbital, c.f. Fig.~\ref{fig-2}(a) and 2(b), becomes resonant with the thermally broadened electrons in the right lead. With increasing $\Delta T$, more of the molecular electron density contributes to the process, balancing ferromagnetic and antiferromagnetic exchanges, which results in a decreased total exchange interaction. The plots in Fig.~\ref{fig-2}(c) shows that we can control this balance into a regime of ferromagnetic exchange for a finite range of temperature biases by tuning the degree of spin-polarization in the leads.

Although previous studies have uncovered that the sign of Heisenberg exchange interaction among magnetic impurities can be tuned electrically (see, e.g., \cite{schwabe1996,diaz2012}), to our knowledge this thermal control of the Heisenberg exchange has never been explored before. More importantly, our general results Eqs.~(\ref{eq-Hs})-(\ref{eq-jmn}) provide a unified microscopic theory for both the electrical and thermal control of magnetic interactions including also anisotropic interactions, as we discuss below.

\begin{figure}[t]
\begin{center}
\includegraphics[width=0.99\columnwidth]{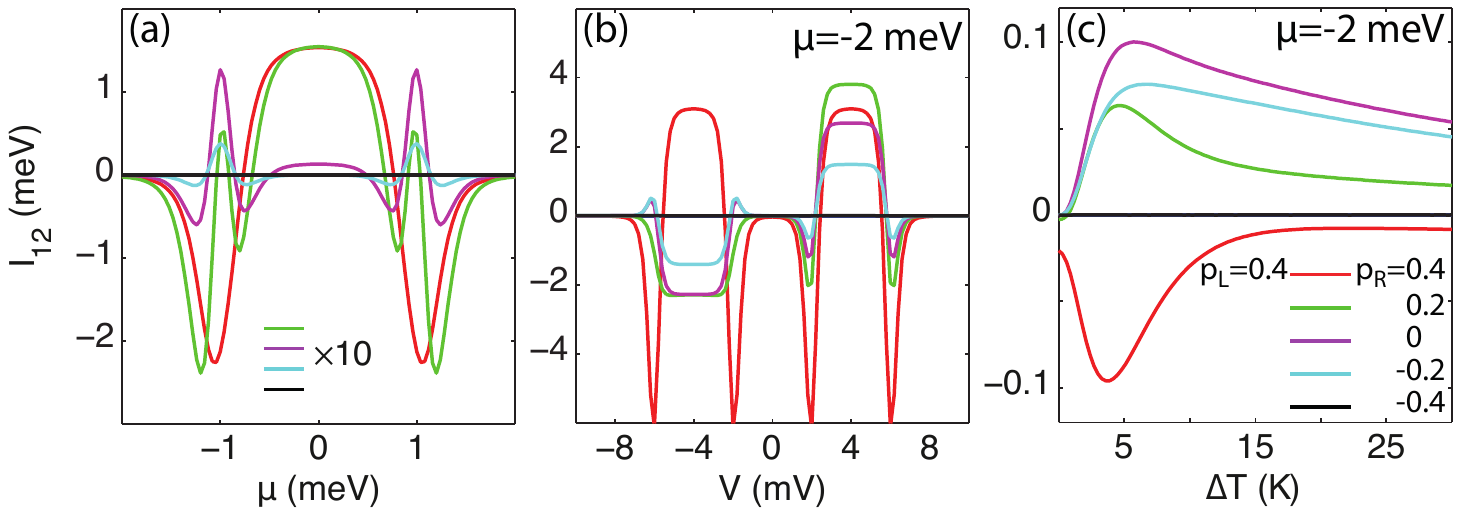}
\end{center}
\caption{(Color online) Ising exchange $I_{12}$. The system is gated ($\mu=-2$ meV) in (b), (c) while other parameters are as in Fig. \ref{fig-2}.}
\label{fig-3}
\end{figure}

Under the same conditions as above, we write the Ising exchange $\bfI_{mn}=I_{mn}\hat{\bf z}\hat{\bf z}$ ($m\neq n$) where
\begin{align}
I_{mn}=&
	-\frac{\mathcal{T}_c^2}{4\pi}
	v_mv_n
	\sum_{\sigma\sigma'}
		\sigma^z_{\sigma\sigma}\sigma^z_{\sigma'\sigma'}
		\Gamma_\sigma
	\fint
		\frac{f_L(\dote{})+f_R(\dote{})}
		{
			|\dote{}-E_{\sigma+}|^2
			|\dote{}-E_{\sigma-}|^2
		}
\nonumber\\&\times
		(\dote{}-\dote{0})		
			\frac{(\dote{}-\dote{0})^2-\mathcal{T}_c^2-(\Gamma_{\sigma'}/4)^2}
			{
				|\dote{}-E_{\sigma'+}|^2
				|\dote{}-E_{\sigma'-}|^2
			}
	d\dote{}.
\label{eq-Idimer}
\end{align}
The basic difference compared to the Heisenberg exchange is that the Ising exchange requires a non-vanishing spin-polarization in the system to be finite. Effectively, the Ising energy becomes a measure of the spin-polarization in the system, which is indicated by the presence of the $z$-component of the Pauli matrices in Eq.~(\ref{eq-Idimer}). Therefore, the Ising energy is small everywhere except when the molecular orbitals are resonant with the chemical potential(s) of the lead(s), see Fig.~\ref{fig-3}. In a similar way as with the Heisenberg energy, we can tune the sign of the Ising exchange by means of gating, voltage bias, thermal bias, and spin-polarization.

\begin{figure}[t]
\begin{center}
\includegraphics[width=0.99\columnwidth]{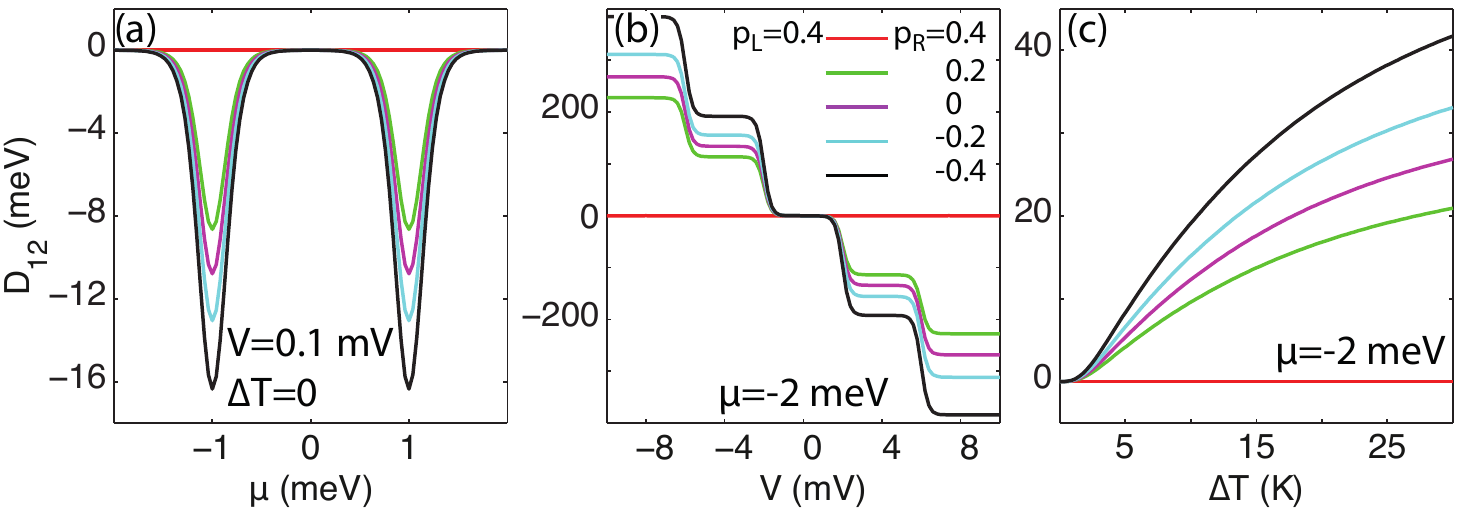}
\end{center}
\caption{(Color online) DM exchange $D_{12}$. The system is biased ($V=0.1$ mV) in (a) and gated ($\mu=-2$ meV) in (b), (c), while other parameters are as in Fig. \ref{fig-2}.}
\label{fig-4}
\end{figure}

Finally the DM exchange energy $\bfD_{mn}=D_{mn}\hat{\bf z}$ ($m\neq n$) within the same approximation but with independent $p_L$ and $p_R$, is obtained as
\begin{align}
D_{mn}=&
	-\frac{1}{\pi}v_mv_n
	\mathcal{T}_c^2(\Gamma_\up^L\Gamma_\down^R-\Gamma_\down^L\Gamma_\up^R)
	\fint
		\Bigl(f_L(\dote{})-f_R(\dote{})\Bigr)
\nonumber\\&\times
		\frac{(\dote{}-\dote{0})^2}
		{
			|\dote{}-E_{\up+}|^2
			|\dote{}-E_{\up-}|^2
			|\dote{}-E_{\down+}|^2
			|\dote{}-E_{\down-}|^2
		}
	d\dote{}.
\label{eq-D12}
\end{align}
The integrand peaks at the resonances $E_{\sigma s}$ while the sign of $D_{mn}$ is governed by the polarities of the voltage bias and temperature difference, and the spin-polarization in the leads. It shows that the DM energy results from the breaking of time-reversal symmetry (spin-polarized current between the localized spins) and space inversion symmetry (biased by a source-drain voltage and/or temperature difference), see Fig.~\ref{fig-4}. The scaling with $\Gamma^2$ suggests that the influence of $\bfD_{mn}$ on the spin excitation spectrum becomes important for stronger coupling $\Gamma$. The combination $\Gamma_\up^L\Gamma_\down^R-\Gamma_\down^L\Gamma_\up^R$, which corresponds to an effective spin-orbit coupling between the leads, suggests that $\bfD_{mn}$ is maximal for antiferromagnetic alignment.

For small voltage bias and zero temperature difference, we have $f_L(\dote{})-f_R(\dote{})\approx eV(\beta/4)\cosh^{-2}[\beta(\dote{}-\mu)/2]$, which indicates a linear voltage bias dependence of $D_{mn}$ near equilibrium, as is shown in Fig.~\ref{fig-4}(b). In case of small temperature difference $\Delta T=T_R-T_L$ and vanishing voltage bias, we have $f_L(\dote{})-f_R(\dote{})\approx-(\Delta T/T)(\beta/4)(\dote{}-\mu)\cosh^{-2}[\beta(\dote{}-\mu)/2]$, indicating a linear dependence on the temperature difference, see Fig.~\ref{fig-4}(c).

The conclusions from the present study of the electrically and thermally mediated exchange interactions between localized magnetic moments have an impact on the magnetic properties of magnetically active quantum devices designed with atomic or molecular building blocks. Depending not only on the couplings to the leads and the spin-polarization in the system but also on gating, voltage bias, and effective temperature difference between the leads, the expected magnetic properties may be drastically different. We expect that our findings should be  verifiable by existing state-of-the-art experiments. We believe that the presented results provide essential new understanding to magnetic interactions and the ability for control by means of external electric and thermal sources.

\acknowledgements
We thank L. Nordstr\"om for stimulating discssions. Support from the Swedish Research Council and Wenner-Gren Foundation (J.F.) is acknowledged. Work at Los Alamos was carried out under the auspices of the National Nuclear Security Administration of the U.S. DOE (Contract No. DE-AC52-06NA25396), and supported by the LANL LDRD Program (J.R.) and by the Center for Integrated Nanotechnologies, a U.S. DOE Office of Basic Energy Sciences user facility (J.-X.Z.).


\begin{thebibliography}{99}
\bibitem{hirjibehedin2006} C. F. Hirjibehedin, C. P. Lutz, and A. J. Heinrich, Science {\bf 312}, 1021 (2006).

\bibitem{wahl2007} P. Wahl, P. Simon, L. Diekh\"oner, V. S. Stepanyuk, P. Bruno, M. A. Schneider, and K. Kern, Phys. Rev. Lett. {\bf 98}, 056601 (2007).

\bibitem{meier2008} F. Meier, L. Zhou, J. Wiebe, and R. Wiesendanger, Science {\bf 320}, 82 (2008).

\bibitem{balashov2009} T. Balashov, T. Schuh, A. F. Tak\'acs, A. Ernst, S. Ostanin, J. Henk, I. Mertig, P. Bruno, T. Miyamachi, S. Suga, and W. Wulfhekel, Phys. Rev. Lett. {\bf 102}, 257203 (2009).

\bibitem{zhou2010} L. Zhou, J. Wiebe, S. Lounis, E. Vedmedenko, F. Meier, S. Bl\"ugel, P. H. Dederichs, and R. Wiesendanger, Nat. Phys. {\bf 6}, 187 (2010).

\bibitem{chen2008} X. Chen, Y.-S. Fu, S.-H. Ji, T. Zhang, P. Cheng, X.-C. Ma, X.-L. Zou, W.-H. Duan, J.-F. Jia, and Q.-K. Xue, Phys. Rev. Lett. {\bf 101}, 197208 (2008).

\bibitem{otte2009} A. F. Otte, M. Ternes, S. Loth, C. P. Lutz, C. F. Hirjibehedin, and A. J. Heinrich, Phys. Rev. Lett. {\bf 103}, 107203 (2009).

\bibitem{pruser2011} H. Pr\"user, M. Wenderoth, P. E. Dargel, A. Weismann, R. Peters, T. Pruschke, and R. G. Ulbrich, Nat. Phys. {\bf 7}, 203 (2011).

\bibitem{khajetoorians2011} A. A. Khajetoorians, J. Wiebe, B. Chilian, and R. Wiesendanger, Science {\bf 332}, 1062 (2011).

\bibitem{loth2012} S. Loth, S. Baumann, C. P. Lutz, D. M. Eigler, and A. J. Heinrich, Science {\bf 335}, 196 (2012).

\bibitem{khajetoorians2013} A. A. Khajetoorians, B. Baxevanis, C. H\"ubner, T. Schlenk, S. Krause, T. O. Wehling, S, Lounis, A. Lichtenstein, D. Pfannkuche, J. Wiebe, and R. Wiesendanger, Science {\bf 339}, 55 (2013).


\bibitem{wende2007} H. Wende, M. Bernien, J. Luo, C. Sorg, N. Ponpandian, J. Kurde, J. Miguel, M. Piantek, X. Xu, PH. Eckhold, W. Kuch, K. Baberschke, P. M. Panchmatia, B. Sanyal, P. M. Oppeneer, and O. Eriksson, Nat. Mat. {\bf 6}, 516 (2007).


\bibitem{fernandez2008} I. Fern\'andez-Torrente, K. J. Franke, and J. I. Pascual, Phys. Rev. Lett. {\bf 101}, 217203 (2008).


\bibitem{chiesa2013} A. Chiesa, S. Carretta, P. Santini, G. Amoretti, and E. Pavarini, Phys. Rev. Lett. {\bf 110}, 157204 (2013).

\bibitem{raman2013} K. V. Raman, A. M. Kamerbeek, A. Mukherjee, N. Atodiresei, T. K. Sen, P. Lazi\'c, V. Caciuc, R. Michel, D. Stalke, S. K. Mandal, S. Bl\"ugel, M. M\"unzenberg, and J. S. Moodera, Nature {\bf 493}, 509 (2013).

\bibitem{fahrendorf2013} S. Fahrendorf, N. Atodiresei, C. Besson, V. Caciuc, F. Matthes, S. Bl\"ugel, P. K\"ogerler, D. E. B\"urgler, and C. M. Schneider, Nat. Commun. {\bf 4}, 2425 (2013).


\bibitem{carretta2004} S. Carretta, P. Santini, G. Amoretti, T. Guidi, R. Caciuffo, A. Candini, A. Cornia, D. Gatteschi, M. Plazanet, and J. A. Stride, Phys. Rev. B {\bf 70}, 214403 (2004).

\bibitem{cornia2004} A. Cornia, A. C. Fabbretti, P. Garrisi, C. Mortal\`{o}, D. Bonacchi, D. Gatteschi, R. Sessoli, L. Sorace, W. Wernsdorfer, and A.L. Barra, Angew. Chem. Int. Ed. {\bf 43}, 1136 (2004).

\bibitem{slageren2002} J. van Slageren, R. Sessoli, D. Gatteschi, A. A. Smith, M. Helliwell, R. E. P. Winpenny, A. Cornia, A. L. Barra, A.G.M. Jansen, E. Rentschler, and G. A. Timco, Chem. Eur. J. {\bf 8}, 277 (2002).

\bibitem{carretta2003} S. Carretta, J. van Slageren, T. Guidi, E. Liviotti, C. Mondelli, D. Rovai, A. Cornia, A.L. Dearden, F. Carsughi, M. Affronte, C.D. Frost, R.E.P. Winpenny, D. Gatteschi, G. Amoretti, and R. Caciuffo, Phys. Rev. B {\bf 67}, 094405 (2003).

\bibitem{troiani2005} F. Troiani, A. Ghirri, M. Affronte, S. Carretta, P. Santini, G. Amoretti, S. Piligkos, G. Timco, and R. E. P. Winpenny, Phys. Rev. Lett. {\bf 94}, 207208 (2005).

\bibitem{carretta2007} S. Carretta, P. Santini, G. Amoretti, T. Guidi, J. R. D. Copley, Y. Qiu, R. Caciuffo, G. Timco, and R. E. P. Winpenny, Phys. Rev. Lett. {\bf 98}, 167401 (2007).

\bibitem{wedge2012} C. J. Wedge, G. A. Timco, E. T. Spielberg, R. E. George, F. Tuna, S. Rigby, E. J. L. McInnes, R. E. P. Winpenny, S. J. Blundell, and A. Ardavan, Phys. Rev. Lett. {\bf 108}, 107204 (2012).

\bibitem{candini2010} A. Candini, G. Lorusso, F. Troiani, A. Ghirri, S. Carretta, P. Santini, G. Amoretti, C. Muryn, F. Tuna, G. Timco, E. J. L. McInnes, R. E. P. Winpenny, W. Wernsdorfer, and M. Affronte, Phys. Rev. Lett. {\bf 104}, 037203 (2010).

\bibitem{santini2011} P. Santini, S. Carretta, F. Troiani, and G. Amoretti, Phys. Rev. Lett. {\bf 107}, 230502 (2011).

\bibitem{mannini2010} M. Mannini, F. Pineider, C. Danieli, F. Totti, L. Sorace, Ph. Sainctavit, M. A. Arrio, E. Otero, L. Joly, J. C. Cezar, A. Cornia, and R. Sessoli, Nature {\bf 468}, 417 (2010). 

\bibitem{mannini2009} M. Mannini, F. Pineider, P. Sainctavit, C. Danieli, E. Otero, C. Sciancalepore, A. M. Talarico, M. A. Arrio, A. Cornia, D. Gatteschi, and R. Sessoli, Nat. Mater. {\bf 8}, 194 (2009).

\bibitem{carretta2006} S. Carretta, P. Santini, G. Amoretti, M. Affronte, A. Candini, A. Ghirri, I. S. Tidmarsh, R. H. Laye, R. Shaw,
and E. J. L. McInnes, Phys. Rev. Lett. {\bf 97}, 207201 (2006). 

\bibitem{leuenberger2001} M. N. Leuenberger and D. Loss, Nature {\bf 410}, 789 (2001).

\bibitem{timco2009} G. A. Timco, S. Carretta, F. Troiani, F. Tuna, R. J. Pritchard, C.A. Muryn, E.J.L. McInnes, A. Ghirri, A. Candini, P. Santini, G. Amoretti, M. Affronte, and R. E. P. Winpenny, Nat. Nanotechnol. {\bf 4}, 173 (2009).

\bibitem{zhang2009} S. Zhang, and S.-L. Zhang, Phys. Rev. Lett. {\bf 102}, 086601 (2009).

\bibitem{bhattacharjee2012} S. Bhattacharjee, L. Nordstr\"om, and J. Fransson, Phys. Rev. Lett. {\bf 108}, 057204 (2012).

\bibitem{fransson2010} J. Fransson, Phys. Rev. B, {\bf 82}, 180411(R) (2010).

\bibitem{szilva2013} A. Szilva, M. Costa, A. Bergman, L. Szunyogh, L. Nordstr\"om, and O. Eriksson, Phys. Rev. Lett. {\bf 111}, 127204 (2013).

\bibitem{misiorny2013} M. Misiorny, M. Hell, and M. R. Wegewijs, Nat. Phys. {\bf 9}, 801 (2013).

\bibitem{note1} In the present paper we disregard external magnetic fields as well as direct (Coloumb induced) exchange interaction of the type $\int\psi^\dagger_\sigma(\bfr)\psi^\dagger_{\sigma'}(\bfr')V(\bfr,\bfr')\psi_\sigma(\bfr')\psi_{\sigma'}(\bfr)d\bfr d\bfr'$, as we want to focus on the interactions that are mediated by the electrons that take part in the conductance.

\bibitem{coronado2004} E. Coronado and P. Day, Chem. Rev. {\bf 104}, 5419 (2004).

\bibitem{zhu2004} J.-X. Zhu, Z. Nussinov, A. Shnirman, and A. V. Balatsky, Phys. Rev. Lett. {\bf 92}, 107001 (2004).
\bibitem{fransson2008} J. Fransson and J.-X. Zhu, New J. Phys. {\bf 10}, 013017 (2008).

\bibitem{IsingE} In case of a spin chiral electronic structure (off-diagonal components of $\bfg^{(1)}_{mn}\neq0$) the Ising interaction also contributes to the transverse anisotropy field. For the self interaction energy this leads to a term like $E(S_x^2+S_y^2)$ in the spin Hamiltonian.

\bibitem{blackman1969} J. A. Blackman and R. J. Elliott, J. Phys. C: Solid State Phys. {\bf 2}, 1670 (1969).

\bibitem{antropov1995} V. P. Antropov, M. I. Katsnelson, M. van Schilfgaarde, and B. N. Harmon, Phys. Rev. Lett. {\bf 75}, 729 (1995).

\bibitem{note2} Note there is an additional factor $1/4$ in our present expression compared to previous results. This is because we consider spin-resolved Green's functions here, and the $\sum_{\sigma}$ can recover a factor of 2. The other factor of 2 can be recovered by considering the unidirectional definition of magnetic interactions in our present work: $J_{mn}$ denotes the exchange interaction from spin $n$ to spin $m$; reciprocally, we have the equivalent $J_{nm}$ from spin $m$ to spin $n$, so that the total exchange interaction between spins $m$ and $n$ is $J_{mn}+J_{nm}=2J_{mn}$.

\bibitem{schwabe1996} N. F. Schwabe, R. J. Elliott, and N. S. Wingreen, Phys. Rev. B {\bf54}, 12953 (1996).
\bibitem{diaz2012} S. D\'{\i}az and \'A. S N\'u\~{n}ez, J. Phys.: Condens. Matter {\bf24}, 116001 (2012).


\end{thebibliography}
\end{document}